\documentclass[prb,
twocolumn,
superscriptaddress,showpacs,amsmath,amssymb]{revtex4}
\usepackage{amsfonts}
\usepackage{bm}
\usepackage{verbatim}

\usepackage{graphicx}

\begin{document}

\title{Two roads to antispacetime in polar distorted B phase: Kibble wall and half-quantum vortex}

 \author{G.E.~Volovik}
\affiliation{Low Temperature Laboratory, Aalto University,  P.O. Box 15100, FI-00076 Aalto, Finland}
\affiliation{Landau Institute for Theoretical Physics, acad. Semyonov av., 1a, 142432,
Chernogolovka, Russia}

\date{\today}

\begin{abstract}
We consider the emergent tetrad gravity and the analog of antispacetime realized 
in the recent experiments\cite{Makinen2019} on the composite defects in superfluid $^3$He:  the  Kibble walls bounded by strings (the half quantum vortices). The antispacetime can be reached in two different ways: by the "safe" route around the Alice string or by dangerous route  across the Kibble wall. This consideration also suggests the scenario of the formation of the discrete symmetry -- the parity $P$ in Dirac equations -- from the continuous symmetry existing on the more fundamental level.
 \end{abstract}
\pacs{
}

\maketitle

\section{Introduction}

The topological materials with emergent analogs of gravity demonstrate the possibility of realization of different exotic spacetimes including the transition to antispacetime, see e.g. Ref. \cite{NissinenVolovik2018} and references therein. There are several routes to the effective gravity. One of them is the tetrad gravity emerging in the vicinity the Weyl or Dirac  points 
\cite{NielsenNinomiya1981,FrogNielBook,Volovik2003,Horava2005}  -- the exceptional crossing points in the fermionic spectrum\cite{Herring1937,Abrikosov1971,Abrikosov1972}. Also the  degenerate $2+1$ gravity  emerges near the Dirac nodal line in the spectrum\cite{NissinenVolovik2018}.
Another important source of gravity is the formation of the tetrads as bilinear combinations of the fermionic fields \cite{Volovik1990,Diakonov2011,Diakonov2012}. 
Also the elasticity tetrads describing the deformation of the crystalline 
materials\cite{DzyaloshinskiiVolovick1980,NissinenVolovik2018b} or quantum vacuum may give rise to gravity.\cite{KlinkhamerVolovik2018}

Different sources of emergent gravity   provide 
different types of the antispacetime obtained by the space reversal $P$ and time reversal $T$ operations, including those where the determinant of the tetrads $e$ changes sign.\cite{Diakonov2011,Diakonov2012,Rovelli2012b,Rovelli2012a,NissinenVolovik2018}
In cosmology,  the antispacetime Universe was in particular suggested as the analytic continuation of our Universe across the Big Bang singularity\cite{Turok2018}. There were the speculations, that the antispacetime may support the nonequilibrium states with negative temperature as a result of analytic continuation across the singularity
\cite{Volovik2019a,Volovik2019b}.

Here we consider the emergent antispacetime realized in the recent experiments 
\cite{Makinen2019}  on the $^3$He analog of the cosmological walls bounded by strings \cite{Kibble1982}, which we call here as Kibble walls. The experiments deal with the time reversal symmetric superfluid phases, where the tetrads   emerge as the  bilinear combinations of the fermionic fields \cite{Volovik1990}.

\section{Relativistic Dirac Green's function}

To see the analogy between relativistic physics and the physics of the Bogoliubov quasiparticles, let us start with the Green's function of the relativistic massive Dirac particle.
In notations \cite{Weinberg1996} used in \cite{Volovik2010},  the Green's function has the form:
 \begin{equation}
S=\frac{Z(p^2)}{-i \gamma^a e_a^\mu p_\mu + M(p^2)}\,.
\label{GreenFunction}
\end{equation}
Here $e_a^\mu$ are tetrads with  $\mu,a=0,1,2,3$; the residue $Z(p^2)$ and the mass $M(p^2)$ are the functions of $p^2=g^{\mu\nu}p_\mu p_\nu$, where
$g^{\mu\nu}=e_a^\mu e_b^\nu \eta^{ab}$.

It is convenient to express $\gamma$-matrices in terms of two sets of Pauli matrices: $\sigma^1$, $\sigma^2$ and $\sigma^3$ for conventional spin,  and $\tau_1$, $\tau_2$, $\tau_3$ for the isospin in the left-right space:
 \begin{equation}
\gamma^0= -i\tau_1 \,\,,\,  \gamma^a= \tau_2\sigma^ a\,\,, \, a=(1,2,3)\,.
\label{gamma0123}
\end{equation}
 \begin{equation}
\gamma_5= -i\gamma^0\gamma^1\gamma^2\gamma^3=\tau_3\,.
\label{gamma5}
\end{equation}

\section{Extension to $^3$He with broken $U(1)$ symmetry}

The time reversal symmetric B-phase and also the polar distorted B-phase (PdB)  of superfluid $^3$He provide the example of the formation 
of the tetrad field as bilinear combination of the fermionic fields 
\cite{Volovik1990,Diakonov2011,Diakonov2012}. 
The Green's function for fermionic Bogoliubov quasiparticles in these superfluids is similar to that in Eq.(\ref{GreenFunction}). Now instead of the mass  function $M(p^2)$, the energy of quasiparticles in the normal Fermi liquid enters, $\epsilon({\bf p})=v_F(|{\bf p}|-p_F)$:
 \begin{equation}
M(p^2) \rightarrow \epsilon({\bf p})\,.
\label{M}
\end{equation}
The $\gamma$-matrices are the same as in Eqs.(\ref{gamma0123}) and (\ref{gamma5}), but their meaning is different. The spin matrices $\sigma^a$ act now in the spin space of $^3$He atoms, while the matrices $\tau_1$, $\tau_2$, $\tau_3$ in the left-right space now correspond to the matrices acting in the isotopic Bogoliubov-Nambu particle-hole space. The function $Z$ is not important, and we ignore it.

The tetrads come from the spin-triplet $p$-wave order parameter in $^3$He superfluids, which is $3\times 3$ matrix $A_a^i$ with spin index $a=(1,2,3)$ and orbital index $i=(1,2,3)$:
 \begin{equation}
\sum_{\bf k} k^i\left<a_{{\bf k}\alpha} a_{-{\bf k}\beta}\right>\sim A_a^i \left( \sigma^a \sigma^2\right)_{\alpha\beta}\,\,, \,  a,i=(1,2,3)\,.
\label{bilinear}
\end{equation}
 For the time reversal symmetric states, the  order parameter has the form:\cite{VollhardtWolfle} 
 \begin{equation}
A_a^i=p_F e^{i\Phi}e_a^i  \,\,, \,  a,i=(1,2,3)\,,
\label{OP}
\end{equation}
(for pure B-phase see \cite{Volovik1990}).

The tetrads $e_a^i $ emerge due to the spontaneously broken symmetries
$SO(3)_S\times SO(3)_L$ under spin and orbital rotations. This is analogous to the formation of the tetrads in relativistic theories as bilinear combinations of the fermionic fields \cite{Diakonov2011,Diakonov2012}. 
However, in addition to tetrads, the  order parameter (\ref{OP}) contains the phase $\Phi$, which numerates the degenerate states obtained after spontaneous breaking of $U(1)$-symmetry in superfluids and superconductors.
The phase $\Phi$ of the order parameter (\ref{OP}) leads to the following modification of the Green's function, which now depends both on the tetrad field $e_a^\mu$ and on the parameter $\Phi$:
 \begin{equation}
\tilde S(e_a^\mu,\Phi)= e^{-\gamma_0 \Phi/2} S (e_a^\mu) e^{\gamma_0 \Phi/2} \,.
\label{GreenFunctionModified}
\end{equation}

%%%%%%%%%%%%%%%%%%%%%%%%%%%%%%%%%%%%%%%%%%%%%%%%%%%%%%%%%
%%%%%%%%%%%%%%%%%%%%%%%%%%%%%%%%%%%%%%%%%%%%%%%%%%%%%%%%%
\begin{figure}[t]
\includegraphics[width=\linewidth]{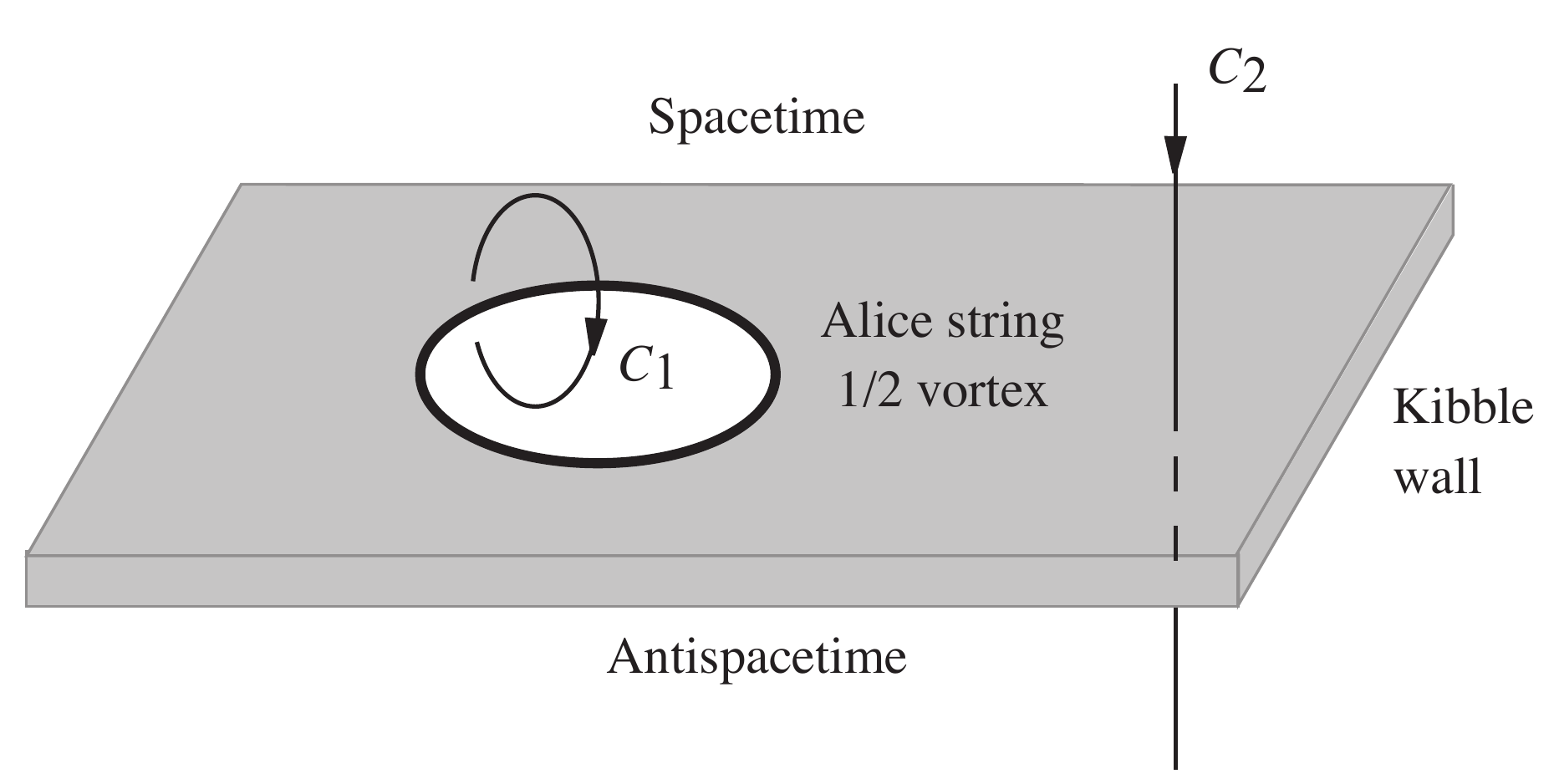}
\caption{ Two roads to antispacetime: the safe route around the Alice string (along the contour $C_1$) or dangerous route along $C_2$ across the Kibble wall (through the Alice looking glass).
}
\label{TwoRoads_Fig}
\end{figure}
%%%%%%%%%%%%%%%%%%%%%%%%%%%%%%%%%%%%%%%%%%%%%%%%%%%%%%%%%
%%%%%%%%%%%%%%%%%%%%%%%%%%%%%%%%%%%%%%%%%%%%%%%%%%%%%%%%%

\section{$U(1)$ symmetry as the origin of parity}

 For $\Phi=\pi$ the symmetry transformation
$e^{-\gamma_0 \Phi/2}$ is equivalent to the conventional space reversal transformation -- the parity  $P=e^{-\gamma_0 \pi/2}=\gamma_0$, with $P^2=-1$.

This suggests that in relativistic theories the discrete symmetry, such as the space inversion $P$, could be the residual $Z_2$ symmetry after breaking of the more fundamental symmetry group at the trans-Planckian level, such as $U(1)$.

On the other hand, the planar phase \cite{VollhardtWolfle}  of superfluid $^3$He  demonstrates the opposite case,  when the discrete $Z_2$ symmetry $C$ of the Hamiltonian becomes continuous  \cite{Makhlin2014}. If the Green's function  commutes  with $C$, the
transformations $e^{i\alpha C}$ generated by the operator $C$  form a
continuous $U(1)$ symmetry group.

\section{Polar distorted B-phase}

In the polar distorted B-phase realized in experiment,\cite{Makinen2019}  the tetrads in the vacuum states are:
 \begin{equation}
e_a^i=c_1\hat {\bf f}_a\hat {\bf x}^ i +c_2\hat {\bf g}_a\hat {\bf y}^ i +  c_3\hat {\bf d}_a\hat {\bf z}^ i
\,\,, \, (a,i)=(1,2,3) \,,
\label{TetradsInhomog}
\end{equation}
where $\hat {\bf d}$, $\hat {\bf f}$ and $\hat {\bf g}$ are orthogonal unit vectors in spin space;
$\hat {\bf x}$, $\hat {\bf y}$ and $\hat {\bf z}$  are orthogonal unit vectors in orbital space; $c_1$, $c_2$ and $c_3$ are the characteristic "speeds of light". In the pure B-phase $|c_1|=|c_2| = |c_3|$, while in the polar phase\cite{Dmitriev2015} one has $c_1=c_2= 0$.

The particular states of the PdB phase are:
 \begin{equation}
e_a^\mu={\rm diag}(-1, c_1,  c_2,  c_3) \,,
\label{TetradsB}
\end{equation}
with $c_2=\pm c_1$ and $|c_2|=|c_1| < |c_3|$.

\section{Kibble wall and Alice string}

The two states with $c_2=+ c_1$ and $c_2=- c_1$ in Eq.(\ref{TetradsB}) can be separated by the nontopological domain wall -- the analog of the Kibble wall bounded by strings \cite{Kibble1982}.
The Kibble walls typically appear in the two phase transitions: at first transition the linear defect (vortex or string) becomes topologically stable; at the second transition the linear defect looses its topological stability  
and becomes the  termination line of the wall -- the Kibble wall. In superfluid $^3$He, the half-quantum vortices appear at first transition from the mormal liquid to the polar phase,\cite{Autti2016} and at further transition to the PdB  phase they become the end lines of the Kibble walls. \cite{Makinen2019}

Across the Kibble wall,  the tetrad  element of  in PdB order parameter, $e^2_2=c_2$, changes sign, and thus  the spacetime transforms to the antispacetime with opposite sign of the tetrad determinant,
${\rm det}\, e$. 
The intermediate state within the Kibble wall has the degenerate tetrad field
\begin{equation}
e_a^\mu={\rm diag}(-1, c_1,  0,  c_3) \,,
\label{DegenerateTetrad}
\end{equation}
and represents the distorted planar phase (in the pure planar phase $|c_1|=|c_3|$ and $c_2=0$ \cite{VollhardtWolfle}). Such types of the nontopological domain walls were originally considered in the B-phase\cite{SalomaaVolovik1988},  and some experimental evidences of them have been recently reported in the thin film of $^3$He-B\cite{Levitin2019}.

 Fig. \ref{TwoRoads_Fig} demonstrates the loop of the half-quantum vortex, which terminates the Kibble wall.  In cosmology, the half-quantum vortex corresponds to the Alice string  \cite{Schwarz1982}. As in the case of the cosmic Alice string, by circling around the half-quantum vortex one continuously arrives at the mirror reflected world. 
Indeed, around the half quantum vortex the phase $\Phi$ changes by $\pi$ and also the vectors 
$\hat {\bf d}$ and $\hat {\bf f}$ rotate by $\pi$. 
As a result, when circling around the HQV, the tetrads are continuously transformed to the state with opposite $e^2_2$:
\begin{equation}
{\rm diag}(-1, c_1,  c_2,  c_3) \rightarrow {\rm diag}(-1, c_1,  -c_2,  c_3) \,,
\label{transformation}
\end{equation}
i.e. to the same antispacetime as across the Kibble wall, but without violation of the PdB state.
The discontinuity in  $e^2_2$ around the Alice string is thus compensated by the Kibble wall, at which $e^2_2$ analytically crosses zero and  changes sign. 
The Kibble wall plays the role of the mirror -- the Alice looking glass. 

\section{Conclusion}

In the polar distorted B-phase of superfluid $^3$He, the half-quantum vortex (Alice string) and the Kibble wall bounded by strings demonstrate the two ways to enter the mirror world in Fig. \ref{TwoRoads_Fig}: either to go around the HQV or to cross the Kibble wall.  The polar distorted B-phase also suggests the scenario of the formation of the discrete symmetry -- the parity $P$ in particle physics -- from the continuous symmetry existing on the more fundamental level.

{\bf Acknowledgements}. 
This work has been supported by the European Research Council (ERC) under the European Union's Horizon 2020 research and innovation programme (Grant Agreement No. 694248).


\begin{thebibliography}{99}


\bibitem{Makinen2019}
 J.T. M\"akinen, V.V. Dmitriev, J. Nissinen, J. Rysti, G.E. Volovik, A.N. Yudin, K. Zhang, V.B. Eltsov,
Half-quantum vortices and walls bounded by strings in the polar-distorted phases of topological superfluid $^3$He,
Nat. Comm. {\bf 10}, 237 (2019),
arXiv:1807.04328.


\bibitem{NissinenVolovik2018}
J. Nissinen and G.E. Volovik,
Dimensional crossover of effective orbital dynamics in polar distorted  $^3$He-A: Transitions to anti-spacetime,
Phys. Rev. D {\bf 97}, 025018  (2018).

 \bibitem{NielsenNinomiya1981}
H.B. Nielsen, M. Ninomiya:
Absence of neutrinos on a lattice.  I - Proof by homotopy theory,
Nucl. Phys. B \textbf{185}, 20  (1981);
Absence of neutrinos on a lattice. II - Intuitive homotopy proof,
Nucl. Phys. B \textbf{193}, 173 (1981).

\bibitem{FrogNielBook}
C.D. Froggatt  and  H.B. Nielsen,
 {\it Origin of Symmetry}, World Scientific, Singapore (1991).

\bibitem{Volovik2003}
G.E. Volovik,
{\it The Universe in a Helium Droplet},
Clarendon Press,  Oxford (2003).

\bibitem{Horava2005}  
P. Ho\v{r}ava,
Stability of Fermi surfaces and $K$-theory,
Phys. Rev. Lett. \textbf{95}, 016405 (2005).

\bibitem{Herring1937}
C. Herring,  
Accidental degeneracy in the energy bands of crystals,
 Phys. Rev. {\bf 52},  365--373  (1937).

\bibitem{Abrikosov1971}
A.A. Abrikosov and S.D. Beneslavskii,
Possible existence of substances intermediate between metals and dielectrics,
JETP {\bf 32}, 699--798 (1971).

\bibitem{Abrikosov1972}
A.A. Abrikosov,
Some properties of gapless semiconductors of the second kind,
J. Low Temp. Phys. {\bf 5}, 141--154 (1972).
 
\bibitem{Volovik1990}
G.E. Volovik, 
Superfluid $^3$He-B and gravity,
 Physica B {\bf 162}, 222-230 (1990).


\bibitem{Diakonov2011}
D. Diakonov,
Towards lattice-regularized Quantum Gravity,
  arXiv:1109.0091.

\bibitem{Diakonov2012}
A.A. Vladimirov, D. Diakonov,
Phase transitions in spinor quantum gravity on a lattice,
Phys. Rev. D {\bf 86}, 104019 (2012). 


\bibitem{DzyaloshinskiiVolovick1980}
I.E. Dzyaloshinskii, and G.E. Volovick,
Poisson brackets in  condensed matter,
Ann. Phys.  {\bf 125}, 67 (1980).

\bibitem{NissinenVolovik2018b}
J. Nissinen and G.E. Volovik,
3+1d QHE, elasticity tetrads and mixed axial-gravitational anomalies,
arXiv:1812.03175.

\bibitem{KlinkhamerVolovik2018}
F.R. Klinkhamer and G.E. Volovik,
Tetrads and $q$-theory,
arXiv:1812.07046.

\bibitem{Rovelli2012b}
M. Christodoulou, A. Riello, C. Rovelli,
How to detect an anti-spacetime,
Int. J. Mod. Phys. D {\bf 21}, 1242014 (2012),
arXiv:1206.3903.

\bibitem{Rovelli2012a}
C. Rovelli, E. Wilson-Ewing,
 Discrete symmetries in covariant LQG,
Phys. Rev. D {\bf 86}, 064002 (2012),
 arXiv:1205.0733.

\bibitem{Turok2018}
L. Boyle, K. Finn and N. Turok,
CPT-Symmetric Universe, 
Phys. Rev. Lett. {\bf 121}, 251301 (2018).

\bibitem{Volovik2019a}
G.E. Volovik,
Negative temperature for negative lapse function,
Pis'ma ZhETF  {\bf 109}, 10-11 (2019),
arXiv:1806.06554.

\bibitem{Volovik2019b}
G.E. Volovik,
Negative temperature in CPT-symmetic Universe,
arXiv:1902.07584.


\bibitem{Kibble1982}
T.W.B. Kibble, G. Lazarides and Q. Shafi,
Walls bounded by strings. 
Phys. Rev. D {\bf 26}, 435--439 (1982).

\bibitem{Weinberg1996}
S. Weinberg, 
The Quantum Theory of Fields,
(Cambridge Univ., Cambridge, 1996),
Section 5.4.

\bibitem{Volovik2010}
G.E. Volovik, 
 Topological invariants  for Standard Model: from semi-metal to topological insulator,
 Pis'ma ZhETF {\bf 91}, 61--67 (2010);   
JETP Lett. {\bf 91}, 55--61 (2010);
arXiv:0912.0502.



\bibitem{VollhardtWolfle}  
D. Vollhardt and  P. W\"olfle 
{\it The superfluid phases of helium 3},  
Taylor and Francis, London (1990).


\bibitem{Makhlin2014}
Yu. Makhlin, M. Silaev, and G.E. Volovik,
Topology of the planar phase of superfluid $^3$He and bulk-boundary correspondence for
three-dimensional topological superconductors,
Phys. Rev. B {\bf 89} 174502 (2014);
arXiv:1312.2677.


\bibitem{Dmitriev2015}
V.V. Dmitriev, A.A. Senin, A.A. Soldatov, and A.N. Yudin,
Polar phase of superfluid  $^3$He in anisotropic aerogel,
Phys. Rev. Lett. {\bf 115}, 165304 (2015).

\bibitem{Autti2016}
S. Autti, V.V. Dmitriev, J.T. M\"akinen, A.A. Soldatov, G.E. Volovik,
A.N. Yudin, V.V. Zavjalov, and V.B. Eltsov,
Observation of half-quantum vortices in superfluid $^3$He,
Phys. Rev. Lett. {\bf 117}, 255301 (2016),
arXiv:1508.02197.


\bibitem{SalomaaVolovik1988}
M.M. Salomaa,  G~E. Volovik, 
Cosmiclike domain walls in superfluid $^3$He-B: Instantons and diabolical points in (${\bf  k}$, ${\bf r}$) space,
Phys. Rev. B {\bf 37}, 9298 - 9311 (1988).


\bibitem{Levitin2019}
L.V. Levitin, B. Yager, L. Sumner, B. Cowan, A.J. Casey, J. Saunders, N. Zhelev, R.G. Bennett, and J.M. Parpia,
Evidence for a spatially modulated superfluid phase of $^3$He under confinement,
Phys. Rev. Lett. {\bf 122}, 085301 (2019).


\bibitem{Schwarz1982}
A. Schwarz, 
Field theories with no local conservation of the electric charge,
Nuclear Physics B{\bf  208}, 141 (1982).




\end{thebibliography}
\end{document}